\documentclass{article}
\usepackage[T1]{fontenc} 
\usepackage[utf8]{inputenc} 
\usepackage{tikz}
\usepackage{adjustbox}
\usetikzlibrary{cd}
\usepackage{microtype}
\usepackage{amsmath, mathtools, amsfonts}
\usepackage{ismir,cite,url}
\usepackage{graphicx}
\usepackage{color}

\title{Cadence Detection in Symbolic Classical Music \\ using Graph Neural Networks}

\oneauthor
  {Emmanouil Karystinaios$^1$ \hspace{1.5cm} Gerhard Widmer$^{1,2}$} {$^1$ Institute of Computational Perception, Johannes Kepler University Linz, Austria \\ $^2$ LIT AI Lab, Linz Institute of Technology, Austria\\ {\tt firstname.lastname@jku.at}}

\def\authorname{E. Karystinaios, G. Widmer}

\begin{document}

\maketitle
\begin{abstract}
Cadences are complex structures that have been driving music from the beginning of contrapuntal polyphony until today. Detecting such structures is vital for numerous MIR tasks such as musicological analysis, key detection, or music segmentation. However, automatic cadence detection remains challenging mainly because it involves a combination of high-level musical elements like harmony, voice leading, and rhythm. In this work, we present a graph representation of symbolic scores as an intermediate means to solve the cadence detection task. We approach cadence detection as an imbalanced node classification problem using a Graph Convolutional Network. We obtain results that are roughly on par with the state of the art, and we present a model capable of making predictions at multiple levels of granularity, from individual notes to beats, thanks to the fine-grained, note-by-note representation. Moreover, our experiments suggest that graph convolution can learn non-local features that assist in cadence detection, freeing us from the need of having to devise specialized features that encode non-local context. We argue that this general approach to modeling musical scores and classification tasks has a number of potential advantages, beyond the specific recognition task presented here.
\end{abstract}

\section{Introduction}\label{sec:introduction}

Graph Neural Networks (GNNs) have recently seen staggering successes in various fields. The MIR community has also experienced the influence of GNNs, principally in the field of recommender systems \cite{korzeniowski2021artist}. However, other sub-branches of MIR could potentially enjoy the graph representation and the benefits of graph deep learning.

Modeling musical scores in all their complexity has been challenging, with many approaches resorting to piano rolls~\cite{huang2019bach}, note arrays~\cite{hawthorne2018enabling}, or custom descriptors~\cite{bigo2018relevance}. In this paper, we present a new representation of the score as a homogeneous graph with note-wise features to model aspects of the score. We use this representation to address the cadence detection task using graph neural networks, treating the task as a node classification problem. More specifically, our contribution is two-fold: a simple graph representation of scores extended with local features, and a Graph Convolutional Network (GCN) model to tackle heavily imbalanced classification tasks such as Cadence Detection.
\textit{Score modeling} itself has two aspects: (1) the construction of the graph, i.e., what are the nodes, and which connections do we define between them; and (2) the choice of score features, and how these relate to their respective graph nodes.
The \textit{classification model} is an adapted version of GraphSMOTE~\cite{zhao2021graphsmote}, a Graph Convolutional Network designed to deal with imbalanced classification problems, which we modified to deal with larger graphs and apply stochastic training. Henceforth, we call this model \textit{Stochastic GraphSMOTE}. 
We employ this model on top of our score modeling with the intention of solving the Cadence Detection task.

The cadence detection setting is binary, i.e., there is a cadence (maybe of a specific type) or not. The current state of the art~\cite{bigo2018relevance} uses an Support Vector Machine (SVM) classifier on a set of custom-designed cadence-specific features, based on three defined "cadence anchor points", and performs score/feature modeling and cadence classification at the level of beats. The model was tested on two annotated datasets: 24 Bach fugues and 42 Haydn string quartet expositions. Our new model proposed here will be shown to achieve comparable overall results; however, we will argue that it makes fewer task-related and musical assumptions, resulting in more general applicability. In particular, our empirical results suggest that by providing local features and applying a Graph Neural Network with neighbor convolution, we can learn nonlocal aspects that help improve prediction. This gives a more general approach for a variety of tasks where features are provided at the level of notes, but prediction may be note-wise, onset-wise, or beat-wise.

The rest of the paper is structured as follows. Section \ref{sec:related_work} discusses related work on cadence detection and music score modeling. Section \ref{sec:modelling} describes the score model and the graph construction from the score, section \ref{sec:problem_setting} introduces the corpora, and section \ref{sec:model} presents the proposed learning algorithm. Section \ref{sec:experiments} presents a series of three experiments and also takes a qualitative look at some examples. Finally, section \ref{sec:conclusion} summarizes and concludes.

\section{Related Work}\label{sec:related_work}




Graphs have emerged as a natural representation of music since the development of Tonnetz by Euler. 
Since then, there have been various proposals to use graph representations for addressing music analysis and MIR tasks. For instance (to name just two), ~\cite{popoff2018relational} introduced relational \textit{Klumpenhouwer networks} for music analysis, and \cite{karystinaios2021music} used \textit{Tonnetz trajectories} for composer classification.
 One can distinguish between \textit{heterogeneous} and \textit{homogeneous} graphs \cite{shi2016survey}. Heterogeneous graphs may have multiple types of edges and nodes, while homogeneous graphs are simpler, containing only a single edge and node type. 
Recently, the creators of VirtuosoNet, a computational model for generating piano performances, used a heterogeneous 
graph representation of the score and trained their system using a Graph Neural Network~\cite{jeong2019graph}. However, in later publications, they reverted to a model without using graphs which achieved better performance~\cite{jeong2019virtuosonet}.
In the present paper, we wish to show that a simple, homogeneous graph representation can form a natural and general basis for modeling a non-trivial music analysis task.


Automatic \textit{cadence detection} is a challenging task. Although cadences are well established concepts, their definition or annotation in music can cause disagreements among musicologists. Previous work on automatic cadence detection has been done by \cite{giraud2015computational} on Bach fugues and by \cite{illescas2007harmonic} for a generalized classical music analysis system. A feature-based approach using standard Machine Learning classifiers is presented in~\cite{bigo2018relevance} which represents the current state of the art. Recently, Sears and Widmer~\cite{sears2021beneath} highlighted the difficulty of detecting textbook voice leading schemata that occur near cadences in written music. However, to our knowledge, there exists no method employing deep learning models to solve the task.

\begin{figure}[b]
    \centering
    \includegraphics[width=0.9\linewidth]{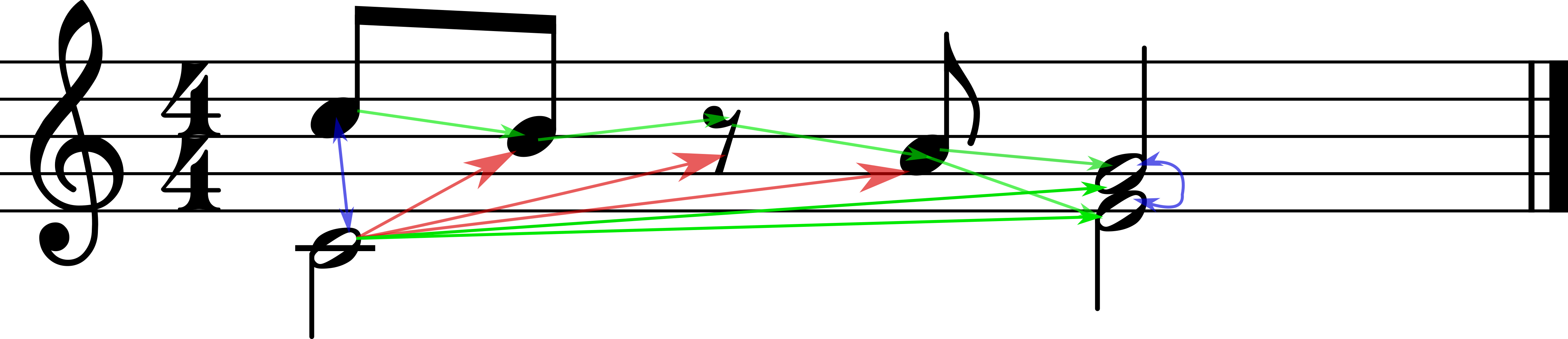}
    \adjustbox{scale=0.8,center}{
    \begin{tikzcd}
    \textrm{note}_a \arrow[r, no head] \arrow[d, no head, bend right]                                       & \textrm{note}_b \arrow[r, no head] \arrow[ld, no head] & \textrm{rest}_k \arrow[r, no head] & \textrm{note}_d \arrow[r, no head] \arrow[rd, no head, bend right] & \textrm{note}_e \arrow[d, no head, bend left] \\
    \textrm{note}_c \arrow[rru, no head] \arrow[rrrr, no head] \arrow[rrru, no head] \arrow[rrrru, no head] &                                                        &                                    &                                                                    & \textrm{note}_f                              
    \end{tikzcd}
    }
    \caption{Example graph creation from a score following the process described in the text.
    $E_{on}$ is denoted in blue, $E_{cons}$ in green, and $E_{dur}$ in red. Global attributes such as time and key signatures are added as node features.}
    \label{fig:my_label}
\end{figure}
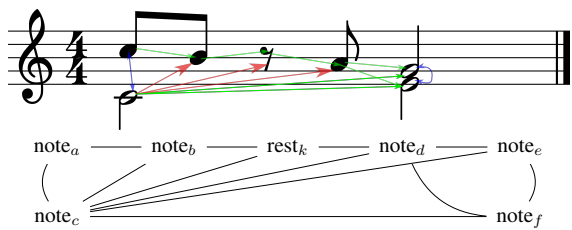

\section{Modeling scores as a graph}\label{sec:modelling}

We model a score as a graph with individual notes and rests as nodes and simple temporal relations as edges. In addition, each graph node is associated with a vector of feature values that represent some basic properties of a note and its immediate context.
Formally, let $G = (V, E)$ be a graph, where $V$ is the set of nodes and $E \subseteq V \times V$ the set of edges and let $A$ be the adjacency matrix of $G$.
Each note and each rest in a score are represented as a node in the graph. We create three types of undirected connections between notes/rests: edges $E_\mathit{on}$ between notes that occur on the same onset; edges $E_\mathit{cons}$ between consecutive notes, and edges $E_\mathit{dur}$ between a note of longer duration and notes whose onsets occur during this time:

\begin{align*}
        E_{on\;} &= \{ (i, j) \mid on(n_i) = on(n_j) \} \\
        E_{cons} &= \{ (i, j) \mid on(n_i) + dur(n_i) = on(n_j) \} \\
        E_{dur\;} &= \{ [ (i, j) \mid  on(n_i) + dur(n_i) > on(n_j) ] \land \\ 
        & \;\;\;\;\;\;\; [ on(n_i) < on(n_j) ] \} \\
        E_{\;\;\;} &= E_{on} \cup E_{cons} \cup E_{dur}
\end{align*}

where $n_i$ is the $i^{th}$ note. $on$ denotes the onset of a note, 
$dur$ the duration. All edges in $E$ are undirected.

\subsection{Feature Overview}\label{subsec:feats}

We use three types of features to further describe a note:\footnote{Code and a complete specification of all features is available on \url{https://github.com/manoskary/cadet}.} general-purpose note-level features to describe a note and its immediate rhythmic/melodic context; general graph topology features to capture aspects of local connectivity; and cadence-specific note features inspired by \cite{bigo2018relevance}.  The third feature category is the only one that is designed with the specific classification target in mind; however, in contrast to \cite{bigo2018relevance}, we restrict these to only consider the immediate local context of a note instead of using positional features relating to predefined past ``cadence anchor points". In this way, we wish to demonstrate the generality of our representation and learning approach, which will hopefully learn more long-distance aspects automatically, as needed.

The first category, \textit{general note-wise features}, is the largest one. For each note in the score, we extract onset time expressed in score-relative beats, duration in beats, and MIDI pitch, using the \textrm{partitura} package \cite{partitura}. Furthermore, we translate global attributes such as time signature and assign them to each note. Also using \textrm{partitura}, we extract a set of generic note-wise features as defined in \cite{chacon2016basis}. Finally, we extract features summarizing intervallic information at the time of onset of each note. These include \textit{interval vectors}\cite{schuijer2008analyzing} and binary features activated when intervallic content is identical to the interval set corresponding to particular chord types, i.e., major, minor, diminished, etc.

Second, we add \textit{graph-aware features} using the first 20 eigenvectors from the Laplacian of the adjacency matrix~\cite{dwivedi2020benchmarking}.

The final category contains \textit{note-wise cadence-related features} similar to those in \cite{bigo2018relevance}, such as voice leading information and voicing. However, our features are calculated at the note level only, considering the time of onset for each note and its immediate neighbors, such as adjacent past onsets or simultaneous onsets. In particular, we do not use any information about events that occur on previous beats. 
While these features are more restricted compared to \cite{bigo2018relevance} they are also more general, since we make no assumptions on and reference to ``cadence anchor points" (e.g., the occurrence of the preceding subdominant and dominant harmony),  which in~\cite{bigo2018relevance} are identified with specialized heuristics.
In total, we store 135 features per node.

\section{Problem Setting \& Corpora}\label{sec:problem_setting}

In this work, we are interested in cadences of the Baroque and Classical periods. The main focus will be on detecting Perfect Authentic Cadences (PAC); where our annotated datasets permit, we will also consider root position Imperfect Authentic Cadences (rIAC) and Half Cadences (HC). The manual annotations in these datasets mark a cadence as occurring on the beat where the final I (i) arrives. Our precise task thus is to predict, for every note of the score, whether this note is contained in a cadence's arrival beat.

To benchmark our method, we used two datasets also used by Bigo et al.~\cite{bigo2018relevance}, and a third one annotated by Allegraud and al.~\cite{allegraud2019learning}. The first set contains the 24 fugues from Bach's Well-tempered Clavier, Book I. The cadence annotations were presented in \cite{giraud2015computational}. The second dataset contains 45 movement expositions from Haydn string quartets; the cadence annotations were produced by Sears and colleagues \cite{sears2018simulating}. The last dataset contains 31 movements of Mozart string quartets with cadence annotations included. All the scores were retrieved from \url{http://kern.ccarh.org} and were parsed in python using the \textrm{partitura} package~\cite{partitura}.\footnote{For reproducibility, we provide the generated graphs that were used for training on \url{https://github.com/manoskary/tonnetzcad}}

\begin{table*}[t]
    \centering
    \begin{small}
    \begin{tabular}{l r r r r r r}
        \textbf{Dataset} & \textbf{Pieces} & \textbf{Nodes} & \textbf{Edges} &
        \textbf{PAC} & \textbf{rIAC} & \textbf{HC} \\
        \hline
         Bach Fugues & 24 & 24,567 & 229,107 & 237 & 78 & 15 \\
         Haydn String Quartets & 45 & 38,661 & 441,491 & 434 & 24 & 340 \\
         Mozart String Quartets & 31 & 68,190 & 762,796 & 1,089 & - & 1,930
    \end{tabular}
    \end{small}
    \caption{Cadence nodes constitute less than 2\% of all nodes.}
    \label{tab:corpus_stats}
\end{table*}

Cadences occur with low frequency in music. In particular, for the corpora we cover in this paper, cadences of all types combined account for less than $2\%$ of the total notes in the score. Our produced score graphs range from approximately $~25$k nodes for the Bach fugues all the way to $~70$k nodes for the Mozart string quartets with more than $750$k edges. Table \ref{tab:corpus_stats} gives detailed dataset statistics. 

\section{Model}\label{sec:model}

\subsection{Graph Convolutional Network}\label{subsec:gcn}

\begin{figure}[!t]
    \centering
    \includegraphics[width=0.8\linewidth]{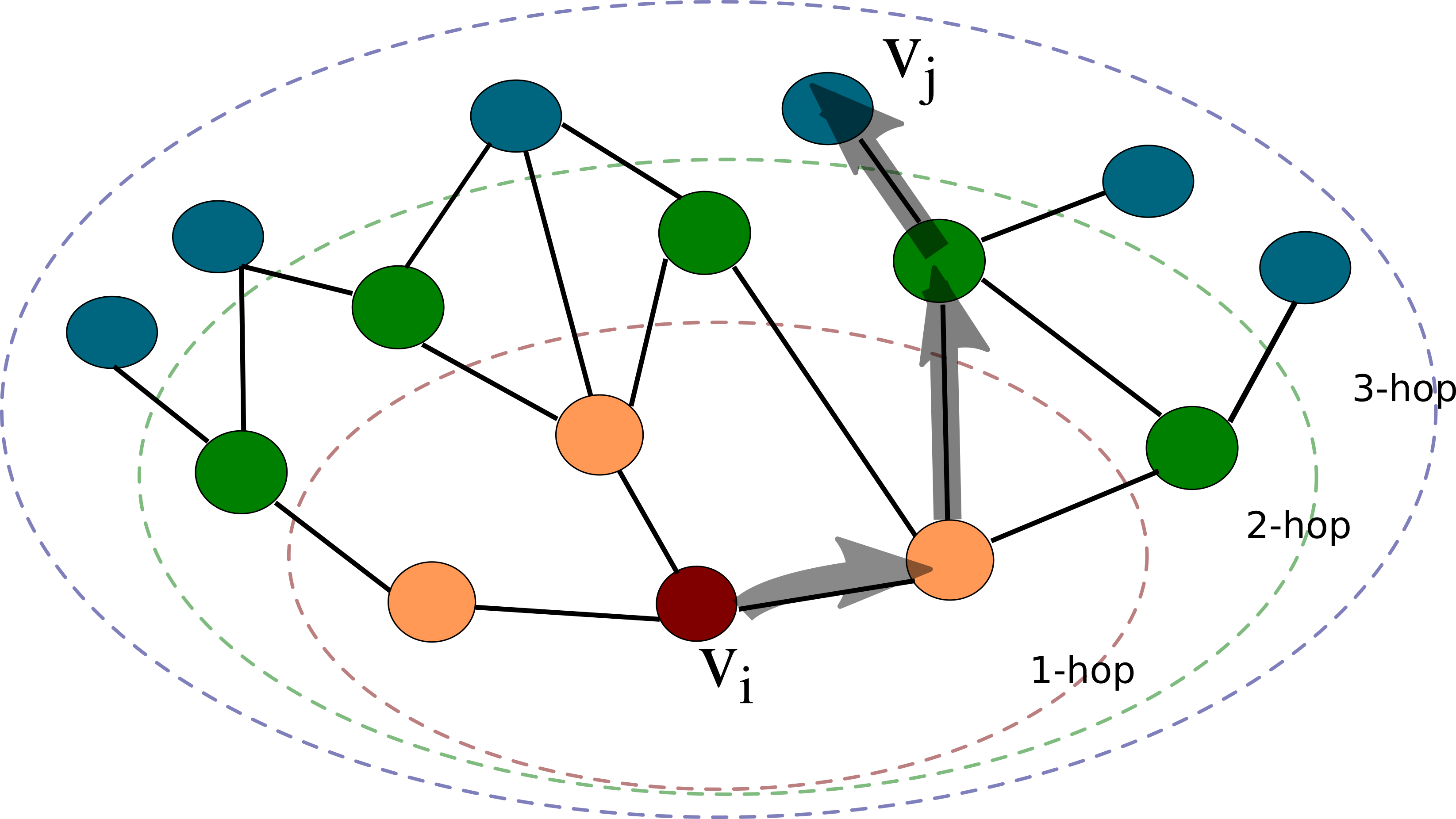}
    \caption{Multi-hop Neighborhood sampling. $v_j$ is 3-hop neighbor of $v_i$. Color cues mark the $k$-hop neighborhoods occurring within the ellipses. The arrows demonstrate a random walk starting from $v_i$ and ending at $v_j$.}
    \label{fig:neigh_sampling}
\end{figure}

The authors of \cite{bigo2018relevance} underline the importance of non-fixed positions for the cadence anchor points. We address this by employing a graph convolutional network. Graph Convolution Networks (GCNs) are based on the same principle as CNNs, but in the context of graphs we encounter the message passing concept, meaning convolution occurs only among nodes connected by edges. This theoretically allows local features to connect with distant features of their $k$-hop neighbors. Therefore, graph representation can learn, using local node information, higher lever information by sampling information from neighbors. Figure~\ref{fig:neigh_sampling} illustrates the neighbor sampling concept.

For our model, we propose \textit{Stochastic GraphSMOTE}, a Graph Convolutional Network with a built-in graph Auto-Encoder and Synthetic Minority Over-sampling for imbalanced node classification. The model consists of 4 parts, the encoder, a SMOTE layer in the encoder's latent space, the decoder, and the classifier. The structure of the model follows GraphSMOTE~\cite{zhao2021graphsmote} but with some major differences, mainly to adapt for stochastic training, which is needed because of the large size of our score graphs.

The encoder applied to a node $i$ is defined as a standard GraphSAGE \cite{hamilton2017inductive} stack given by:
\begin{align*}
        \mathbf{h}_{\mathcal{N}(i)}^{(l+1)} &= \mathrm{mean}\left(\{\mathbf{W}_{pool}^{(l+1)} \cdot \mathbf{h}_{j}^{l}, \forall j \in \mathcal{N}(i) \}\right)\\
        \mathbf{h}_{i}^{(l+1)} &= \sigma \left(\mathbf{W}_{enc}^{(l+1)} \cdot \mathrm{concat}(\mathbf{h}_{i}^{l}, \mathbf{h}_{\mathcal{N}(i)}^{l+1}) \right) \\
        \mathbf{h}_{i}^{(l+1)} &= \mathrm{norm}(\mathbf{h}_{i}^{l})
\end{align*}

where $h_i^{(l)}$ is the hidden representation of node $i$ on layer $l$, $\sigma$ is an activation function, $\mathrm{norm}$ is a normalization function, $\mathbf{W}$ are learnable weights, and $\mathcal{N}(i) = \{j \mid (i, j) \in E\}$ are the neighbors of node $i$. Let $B \subseteq V$ a subset of nodes denoting a batch sample. Then, given $L$ the total number of hidden layers, $\mathbf{H}^{(enc)}_B = \set{\mathbf{h}_u^{(L+1)} \mid u \in B}$.

\subsection{Dealing with Extreme Class Imbalance: Stochastic GraphSMOTE}

Since cadences are very sparse, we need to introduce a balancing technique in order to avoid gradient convergence that will result in predicting only the majority class, i.e., absence of cadence. To counter this effect, we introduce a SMOTE layer that is applied in the \textit{latent space} of the encoder. SMOTE generates synthetic samples with the same label as the minority class (see \cite{chawla2002smote} for details). The main novelty of our model is that the SMOTE is performed for each batch separately.

In each batch, we count the occurrence, $\mu_i$, for each of the classes $i \in I$. In the binary setting, let $\mu_M$ be the number of samples with the same label as the majority class and $\mu_m$ be the number of samples with the same labels as the minority class. By generating $(\mu_M - \mu_m)$ samples with the same label as the minority class, we force a $1:1$ binary class distribution. To generate these samples, in each batch, we randomly select a sample instance of the minority class as an anchor point and gather the $k$ nearest neighbor samples of the same class within the batch. Finally, $\mu$ samples are generated as random linear interpolations between a randomly selected neighbor out of the $k$, and the selected anchor point in the euclidean space. 
Performing SMOTE in the latent space assumes that a more appropriate representation for the generation of the synthetic minority samples is learned.

If $\mathbf{H}^{(enc)}_B$ is the hidden representation of the batch sampled nodes after the encoder layer, then $\mathbf{H}^{(smote)}_B$ is the SMOTE upsampling algorithm applied on $\mathbf{H}^{(enc)}_B$. Our Decoder layer is responsible for generating edges within the original nodes of the graph and the synthetic ones, created by SMOTE. The decoder output is described by the following equation:

\begin{align*}
        \mathbf{A}_B^{(dec)} &= \sigma \left( \mathbf{H}^{(smote)}_B  \cdot \mathbf{W}^{(dec)}  \cdot \mathrm{transpose}(\mathbf{H}^{(smote)}_B) \right) \\
        \mathbf{A}_B^{(thr)} &= \mathrm{hardshrink}\left(\mathbf{A}^{(dec)}_B, \tau \right)
\end{align*}

where $W^{(dec)}$ are the decoder's learnable weights, $\sigma$ is a sigmoid activation function,  and $\mathrm{hardshrink}$ is the hard shrinkage function with threshold $\tau$. $\mathbf{A}^{(dec)}_B$ is the generated adjacency from the decoder and $\mathbf{A}_B^{(thr)}$ is a thresholded adjacency by a factor $\tau$.

We define a regularization loss that aims at constraining the generated adjacency close to the original, defined by:

\begin{equation*}
    \mathcal{L}_B^{(dec)} = \mathrm{BCE}\left(\mathbf{A}_B^{(dec)}, \mathbf{A}_B\right)
\end{equation*}

where BCE is the binary cross entropy loss, $\mathbf{A}_B^{(dec)}$ is the generated adjacency of the decoder for batch sample $B$ and $A_B$ is the adjacency matrix for batch sample $B$. Since we learn an edge generator which is good at reconstructing the adjacency matrix using the encoder's latent representations, it should also give adequate edge predictions for synthetic nodes.

The GNN classifier is composed of a GraphSAGE layer~\cite{hamilton2017inductive} with a linear layer on top. By adding a graph convolution layer such as GraphSAGE in the classifier, we can benefit from learning information from the generated adjacency and the neighbors of nodes. The GraphSAGE layer of the classifier is slightly different from the encoder because it performs directly on the generated thresholded adjacency of each batch sample:

\begin{align*}
        \mathbf{h}_{\mathcal{N}(i)}^{(clf)} &= \textrm{mean}\left( \mathbf{W}^{(pool)} \cdot \mathbf{A}^{(thr)}_B[i, :] \cdot \mathbf{H}^{(enc)}_B \right) \\
        \mathbf{h}_{i}^{(clf)} &= \textrm{norm}\left( \sigma \left( \mathbf{W}^{(clf)} \cdot \mathrm{concat} \left( \mathbf{h}^{(enc)}_i, \mathbf{h}^{(clf)}_{\mathcal{N}(i)} \right) \right) \right) \\
        \mathbf{h}_{i}^{(clf)} &= \mathrm{softmax}(\mathbf{W}^{(proj)} \cdot \mathbf{h}_{i}^{(clf)})
\end{align*}
where $\mathbf{h}_{i}^{(clf)}$ are the predicted class probabilities of node $i$, $\textbf{W}$ are learnable weights, $\mathbf{A}_B^{(thr)}$ is the generated thresholded adjacency from the decoder, $\mathbf{H}_B^{(enc)}$ are the batch encodings of the encoder and $\mathbf{h}^{(enc)}_i$ is the encoder's output for node $i$. During training, we use $\mathbf{H}^{(smote)}_B$ and $\mathbf{h}^{(smote)}_i$ respectively instead of $\mathbf{H}_B^{(enc)}$ and $\mathbf{h}^{(enc)}_i$. 
We define the \textit{total loss} of our model for batch samples $B$:

\begin{equation*}
    \mathcal{L}^{(tot)}_B = \mathcal{L}^{(CE)}_B + \gamma * \mathcal{L}^{(dec)}_B
\end{equation*}
where $\mathcal{L}_{CE}$ signifies the cross entropy loss and $\gamma$ is a hyper parameter.

\begin{table*}[t]
    \centering
    \begin{tabular}{l||l r r r r r}
        \hline
        \textbf{Dataset} & \textbf{Model}  & \textbf{F1 Note} & \textbf{F1 Onset} & \textbf{F1 Beat}  & \textbf{Prec. Beat} & \textbf{Recall Beat}\\
        \hline \hline
         & Bigo et al.~model   & - & - & \textbf{0.80} & \textbf{0.89} & 0.72 \\
        Bach Fugues (PAC) & SGSMOTE& 0.85 & 0.75 & 0.73 & 0.70 & 0.77 \\
        \small{(12 fugues)} & Pretrained SGSMOTE & \textbf{0.90} & \textbf{0.83}  & \textbf{0.80} & 0.74 & \textbf{0.89} \\
        \hline
         & Bigo et al.~model  & - & - & 0.68 & 0.71 & 0.65  \\
        Bach Fugues (rIAC) & SGSMOTE& \textbf{0.87} & \textbf{0.75} & \textbf{0.73} & \textbf{0.75} & 0.72  \\
         \small{(12 fugues)}& Pretrained SGSMOTE & 0.87 & 0.73 & 0.71 & 0.62 & \textbf{0.82}  \\
        \hline
        & Bigo et al.~model & - & - &\textbf{0.69} & \textbf{0.60} & \textbf{0.82}   \\
        Haydn String Quartets (PAC) & SGSMOTE & 0.77 & 0.56 & 0.59 & 0.47 & 0.78  \\
        \small{(21 pieces)} & Pretrained SGSMOTE & \textbf{0.81} & \textbf{0.63} & 0.64 & 0.54 & 0.78  \\
        \hline
        & Bigo et al.~model  & - & - & 0.29 & 0.19 & \textbf{0.56}  \\
        Haydn String Quartets (HC) & SGSMOTE & 0.65 & 0.32 & 0.30 & 0.33 & 0.27  \\
        \small{(21 pieces)} & Pretrained SGSMOTE & \textbf{0.69} & \textbf{0.44}  & \textbf{0.41}& \textbf{0.41} & 0.41 \\
        \hline
    \end{tabular}
    \caption{Results using half of the dataset for training, half for testing. Bach: fugues no.1-12 were used for training, no.13-24 for testing; Haydn: random 21:21 split.
    The pretrained network was trained on the other dataset, i.e. \textit{Pretrained SGSMOTE} for Bach Fugues was pre-trained on string quartets, etc. Classification is binary, the presented $F1$ scores are for the positive class, i.e., the cadence (PAC: Perfect Authentic Cadence; rIAC: root position Imperfect AC; HC: Half Cadence).}
    \label{tab:sota_results}
\end{table*}

Our model is trained stochastically, meaning that to create each batch a subset $B$ of nodes are sampled. From these sampled nodes, given a pre-defined depth $k$, we retrieve the immediate neighbors of every $v \in B$ up to their $k$-hop neighbors in the graph $G$. We use neighbor sampling to reduce the cost of retrieving all up to $k$-hop neighbors of $v$ by defining a maximum number $\phi_l$ of neighbors per depth layer $l$.

\section{Experiments}\label{sec:experiments}

We conduct three main experiments. The first compares our model to the state of the art results in \cite{bigo2018relevance}, using the same data and train/test setup. The second experiment focuses on multi-class learning of the particular type of cadence using different sets of features, in order to investigate how the model generalizes to a more complex setting and inspect the relevance of different feature sets. The third experiment investigates how neighbor convolution contributes to the model's performance.\footnote{All results, experiments, and the trained models are available on \url{https://wandb.ai/melkisedeath/Cadence Detection}
}

We fix our model with a hidden dimension of 256, with $L=2$ hidden layers with $\phi_1 = 10$ and $\phi_2 = 25$ sampled neighbors for hidden layers 1 and 2 of the encoder, respectively, and one hidden layer of the same dimension for the classifier. The learning rate is set at 0.007, the weight-decay at 0.007, with a batch size of 1024, $k=3$ for SMOTE, the decoder regularization loss multiplier $\gamma=0.5$, and adjacency threshold value $\tau=0.5$.

\begin{figure*}
    \centering
    \includegraphics[width=\textwidth]{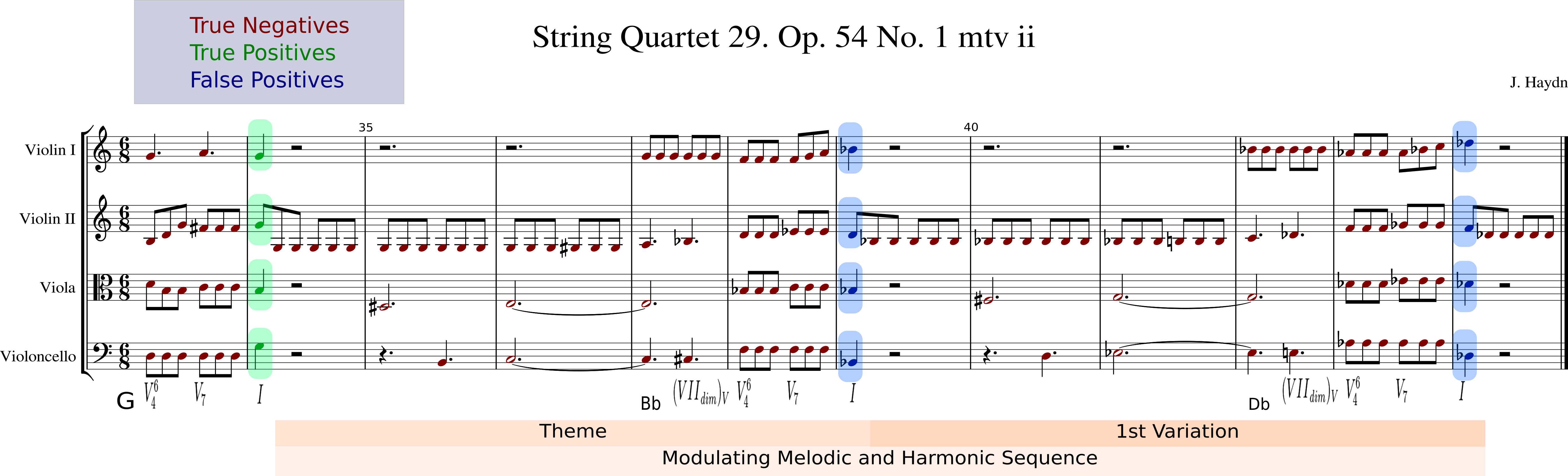}
    \caption{Haydn's String Quartet 29. Op.54 No.1 Mvt. II, mm.~33-45. Showing the output of the Stochastic GraphSMOTE Network for PAC prediction. True negatives are marked with red, true positives with green, false positives with blue. A partial analysis shows the chords towards the end of cadences and highlights a modulating sequence where every sequence ends with a cadential pattern, which counts as false positive predictions by the network. }
    \label{fig:hsq_predictions}
\end{figure*}

\subsection{Quantitative Results}

Table \ref{tab:sota_results} summarizes the results of the first experiment, comparing our model's performance to the state of the art.\footnote{In accordance with \cite{bigo2018relevance}, we ignore the HC in Bach and rIAC in Haydn, because of their low numbers.} The reference model \cite{bigo2018relevance} can only classify at the beat level; our representation and classification model are more flexible in this regard, as they have access to, and describe, individual notes. In particular, our model can provide predictions at three different levels, note-wise, onset-wise and beat-wise predictions (the latter two simply by aggregation). In Table~\ref{tab:sota_results} we present the results of these predictions at all levels, in terms of F1 score.
Only beat-wise scores are given for the reference model (taken from~\cite{bigo2018relevance}). The last two columns of table~\ref{tab:sota_results} give the recall and precision for the beat-wise prediction. All metrics are presented for the positive, i.e. minority/cadence, class.

Our model matches or slightly surpasses the state of the art in rIAC detection in Bach fugues and on HCs in Haydn string quartets but does not reach the reference model's F1 results in PAC detection. We additionally present a pre-trained version of Stochastic GraphSMOTE, where the network was first trained on additional data and fine-tuned for the task. Specifically, the network for PAC prediction in Bach was pre-trained on the string quartets and vice versa. Pre-training, and thus the need for additional data, is the price we pay for the generality of the graph representation and the consequent size (number of parameters) of the deep network.
Pre-training helps to (markedly) improve the results on HC, catch up with the reference on PAC in Bach, and narrow the gap on PAC in Haydn.

Generally, our results agree with \cite{bigo2018relevance} in implying that half cadences (HC) seem significantly harder to identify than authentic cadences, both perfect and imperfect.
Another, more specific, observation concerns different ways in which the compared models achieve their overall F1 scores. In the PAC detection tasks, in particular, we observe comparable or higher recall of our model compared to the reference, but lower precision. This observation motivated us to check some of our model's false positive predictions; Section \ref{subsec:qualitative} below will show several instructive examples of `almost correct' identifications.

The \textit{second experiment} we conducted (Table \ref{tab:multiclass_results}) focuses on comparing the relevance of feature groups. For compactness we present here a multi-class classification scenario where we account not only for the existence of a cadence but also for the type of cadence present; that is, we have tree-class problems: no cadence, PAC, or rIAC (Bach) / HC (Haydn, Mozart).
We compare two configurations: using all available features (as in the first experiment, feature set \textit{all} in the table), or only feature sets 1 and 2, excluding the cadence-specific features (category 3 in Section~\ref{subsec:feats}; marked \textit{general} in the table). Given this 3-class setting, we chose to report 
the macro averaged F1 score over all three classes. (Macro averaging was chosen to counter the overwhelming effect of the majority class \textit{no cadence}). The results (see Table \ref{tab:multiclass_results}) support the relevance of carefully devised cadence-related features \`a la \cite{bigo2018relevance}. However, also the general-purpose category 1 and 2 features alone support non-trivial cadence recognition and discrimination performance, which implies that the relational graph representation in combination with a convolutional approach manages to enrich highly local features with relevant non-local score context.

\begin{table}[t]
    \begin{tabular}{l || r r r}
        \hline
        \textbf{Dataset} & \textbf{Features} & \textbf{F1 Note} & \textbf{F1 Beat}\\
        \hline \hline
        Bach Fugues & general & 0.602 & 0.667  \\
        (PAC \& rIAC) & all & 0.653 & 0.702\\
         \hline
        Haydn String Quart. & general & 0.542 & 0.610\\
        (PAC \& HC) & all & 0.648 & 0.663\\
        \hline
        Mozart String Quart. & general & 0.584 & 0.569 \\
        (PAC \& HC) & all & 0.588 & 0.606 \\
        \hline
    \end{tabular}
    \caption{Three-class cadence classification with two different feature sets. Results were obtained by 5 fold cross validation (70\% of pieces for training, 10\% validation, 20\% testing); no pre-training. 
    Feature set \textit{all} contains all features from Section \ref{subsec:feats}; \textit{general} excludes Category 3 cadence-specific engineered features.
    }
    \label{tab:multiclass_results}
\end{table}

To investigate this latter aspect in more detail, we run a \textit{third experiment}, to look at the effect of neighbor convolution depth on the obtainable classification score, again at three prediction granularity levels (note, onset, beat). Convolution depth refers to the number $l$ of hidden layers of the encoder and the subsequent neighbor sampling up to $l$-hop neighbors.
Our results (see Table~\ref{tab:depth}) suggest that neighbor convolution clearly contributes to learning non-local features. Best results are achieved when using a convolution depth of 2. Increasing the receptive field beyond that level, we observed some instabilities emerging in the learning model, which could be attributed to the common vanishing gradient problem in deep GCNs~\cite{li2019deepgcns}.

\begin{table}[!h]
    \centering
    \begin{tabular}{l r r r}
        \hline
        \textbf{Depth} & \textbf{F1 Note} & \textbf{F1 Onset} & \textbf{F1 Beat}\\
        \hline
        None & 0.833 & 0.671 & 0.667 \\
        1-hop & 0.854 & 0.707 & 0.701 \\ 
        2-hop & \textbf{0.869} & \textbf{0.737} & \textbf{0.732} \\
        3-hop & 0.836 & 0.706 & 0.659 \\
        \hline
    \end{tabular}
    \caption{Effect of neighbor convolution depth on PAC prediction in Bach fugues. The F1 Note/Onset/Beat scores presented are binary, i.e., for the PAC class. Depth refers to neighbor convolution depth. \textit{None} means no graph convolution.
    }
    \label{tab:depth}
\end{table}

\subsection{A Qualitative Look}\label{subsec:qualitative}

Motivated by the fact that our model, while higher on recall, seems to be lower on precision than the model in~\cite{bigo2018relevance}, we take a closer look at some of the false positives in individual examples. Our findings suggest that many false positive predictions resemble cadences, in terms of tonal structure or implications, and could be considered and annotated as such, but lack some main components.

\begin{figure}[t]
    \centering
    \includegraphics[width=\linewidth]{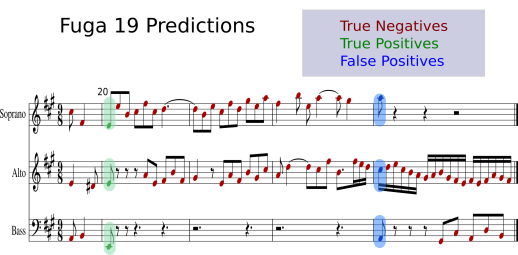}
    \caption{Predictions of Stochastic GraphSMOTE for fugue No.19, J.S.Bach, Well-tempered Clavier.}
    \label{fig:cad_predictions}
\end{figure}

Figure~\ref{fig:cad_predictions} shows an example. The cadence prediction by our model on the downbeat of bar 23 is a false positive, according to the ground truth annotation. However, one could argue that the passage clearly has a cadence-like role, marking the end of the $2^{\text{nd}}$ fugal episode and the return to the original tonality of A major \cite{tonic_chord_2018}.

Another example is the passage discussed in Fig.4 of \cite{bigo2018relevance}, where a pattern occurs that has all the technical ingredients of a PAC, but was not annotated as such for (debatable) higher-level musicological considerations. Again, our model's PAC prediction there counts as a false positive.

As a final example, consider mm.~33-45 of Haydn's Op.54 No.1, $2^{\text{nd}}$ mvt (Figure ~\ref{fig:hsq_predictions}). We observe two false positive beat-wise predictions (8 if we count note-wise) in bars 39 and 44, respectively, following a true PAC on the beginning of bar 34. A harmonic analysis of these bars indicates a proper PAC preparation with text-book voice leading on the cadence arrival point in every occasion. These two false positive PACs form part of a modulating melodic and harmonic sequence; whether to classify them as cadences is a matter of higher-level musicological considerations.

We cite these few qualitative examples in an attempt to show that our prediction model can identify many more cadential patterns than the raw experimental figures suggest, but by design cannot consider high-level musical considerations such as, e.g., whether PAC-like patterns that occur in sequence should count as PACs or not.

\section{Conclusion}\label{sec:conclusion}

We have presented a graph approach to effectively target the cadence detection task on symbolic classical scores. We demonstrated that our Graph Convolutional Network, Stochastic GraphSMOTE, can learn using only local note features, without the need for any musical assumptions about cadence anchor points.
Furthermore, our network can produce fine-grained predictions at the level of individual notes.

Future work will address the performance of the model on different tasks, using the same graph representation. We hope to be able to show that this simple but general and natural representation of scores in terms of graphs can support a broad variety of symbolic music analysis and classification tasks.

\section{Acknowledgements}

This work is supported by the European Research Council (ERC) under the EU’s Horizon 2020 research \& innovation programme, grant agreement No.~101019375 (“Whither Music?”), and the Federal State of Upper Austria (LIT AI Lab). The authors would like to thank Dr.~Hamid Eghbal-zadeh for helpful discussions on Graph Neural Networks.

\bibliography{ISMIRtemplate}

\begin{thebibliography}{10}
\providecommand{\url}[1]{#1}
\csname url@samestyle\endcsname
\providecommand{\newblock}{\relax}
\providecommand{\bibinfo}[2]{#2}
\providecommand{\BIBentrySTDinterwordspacing}{\spaceskip=0pt\relax}
\providecommand{\BIBentryALTinterwordstretchfactor}{4}
\providecommand{\BIBentryALTinterwordspacing}{\spaceskip=\fontdimen2\font plus
\BIBentryALTinterwordstretchfactor\fontdimen3\font minus
  \fontdimen4\font\relax}
\providecommand{\BIBforeignlanguage}[2]{{%
\expandafter\ifx\csname l@#1\endcsname\relax
\typeout{** WARNING: IEEEtran.bst: No hyphenation pattern has been}%
\typeout{** loaded for the language `#1'. Using the pattern for}%
\typeout{** the default language instead.}%
\else
\language=\csname l@#1\endcsname
\fi
#2}}
\providecommand{\BIBdecl}{\relax}
\BIBdecl

\bibitem{korzeniowski2021artist}
F.~Korzeniowski, S.~Oramas, and F.~Gouyon, ``Artist similarity with graph
  neural networks,'' in \emph{Proceedings of the 22nd International Society for
  Music Information Retrieval Conference}, 2021.

\bibitem{huang2019bach}
C.-Z.~A. Huang, C.~Hawthorne, A.~Roberts, M.~Dinculescu, J.~Wexler, L.~Hong,
  and J.~Howcroft, ``The bach doodle: Approachable music composition with
  machine learning at scale,'' in \emph{Proceedings of the 18th International
  Society for Music Information Retrieval Conference}, 2019.

\bibitem{hawthorne2018enabling}
C.~Hawthorne, A.~Stasyuk, A.~Roberts, I.~Simon, C.-Z.~A. Huang, S.~Dieleman,
  E.~Elsen, J.~Engel, and D.~Eck, ``Enabling factorized piano music modeling
  and generation with the maestro dataset,'' in \emph{Proceedings of 7th
  International Conference on Learning Representations}, 2019.

\bibitem{bigo2018relevance}
L.~Bigo, L.~Feisthauer, M.~Giraud, and F.~Lev{\'e}, ``Relevance of musical
  features for cadence detection,'' in \emph{Proceedings of the 19th
  International Society for Music Information Retrieval Conference (ISMIR
  2018)}, 2018.

\bibitem{zhao2021graphsmote}
T.~Zhao, X.~Zhang, and S.~Wang, ``Graphsmote: Imbalanced node classification on
  graphs with graph neural networks,'' in \emph{Proceedings of the 14th ACM
  International Conference on Web Search and Data Mining}, 2021.

\bibitem{popoff2018relational}
A.~Popoff, M.~Andreatta, and A.~Ehresmann, ``Relational poly-klumpenhouwer
  networks for transformational and voice-leading analysis,'' \emph{Journal of
  Mathematics and Music}, vol.~12, no.~1, 2018.

\bibitem{karystinaios2021music}
E.~Karystinaios, C.~Guichaoua, M.~Andreatta, L.~Bigo, and I.~Bloch, ``Music
  genre descriptor for classification based on tonnetz trajectories,'' in
  \emph{Proceedings of Journ{\'e}es Informatiques Musicales}, 2021.

\bibitem{shi2016survey}
C.~Shi, Y.~Li, J.~Zhang, Y.~Sun, and S.~Y. Philip, ``A survey of heterogeneous
  information network analysis,'' \emph{IEEE Transactions on Knowledge and Data
  Engineering}, vol.~29, no.~1, pp. 17--37, 2016.

\bibitem{jeong2019graph}
D.~Jeong, T.~Kwon, Y.~Kim, and J.~Nam, ``Graph neural network for music score
  data and modeling expressive piano performance,'' in \emph{International
  Conference on Machine Learning}, 2019.

\bibitem{jeong2019virtuosonet}
D.~Jeong, T.~Kwon, Y.~Kim, K.~Lee, and J.~Nam, ``Virtuosonet: A hierarchical
  rnn-based system for modeling expressive piano performance.'' in
  \emph{Proceedings of the 20th International Society of Music Information
  Retrieval Conference}, 2019.

\bibitem{giraud2015computational}
M.~Giraud, R.~Groult, E.~Leguy, and F.~Lev{\'e}, ``Computational fugue
  analysis,'' \emph{Computer Music Journal}, vol.~39, no.~2, 2015.

\bibitem{illescas2007harmonic}
P.~R. Illescas, D.~Rizo, and J.~M.~I. Quereda, ``Harmonic, melodic, and
  functional automatic analysis,'' in \emph{Proceedings of the International
  Computer Music Conference}, 2007.

\bibitem{sears2021beneath}
D.~R. Sears and G.~Widmer, ``Beneath (or beyond) the surface: Discovering
  voice-leading patterns with skip-grams,'' \emph{Journal of Mathematics and
  Music}, vol.~15, no.~3, 2021.

\bibitem{partitura}
C.~E.~C. Chac{\'o}n, P.~Silvan, E.~Karystinaios, F.~Foscarin, M.~Grachten, and
  G.~Widmer, ``Partitura: A python package for symbolic music processing,'' in
  \emph{Proceedings of the Music Encoding Conference}, 2022.

\bibitem{chacon2016basis}
C.~E.~C. Chac{\'o}n, ``Computational modeling of expressive music performance
  with linear and non-linear basis function models,'' Ph.D. dissertation,
  Johannes Kepler University, Austria, 2018.

\bibitem{schuijer2008analyzing}
M.~Schuijer, \emph{Analyzing atonal music: Pitch-class set theory and its
  contexts}.\hskip 1em plus 0.5em minus 0.4em\relax University Rochester Press,
  2008.

\bibitem{dwivedi2020benchmarking}
V.~P. Dwivedi, C.~K. Joshi, T.~Laurent, Y.~Bengio, and X.~Bresson,
  ``Benchmarking graph neural networks,'' \emph{arXiv preprint
  arXiv:2003.00982}, 2020.

\bibitem{allegraud2019learning}
P.~Allegraud, L.~Bigo, L.~Feisthauer, M.~Giraud, R.~Groult, E.~Leguy, and
  F.~Lev{\'e}, ``Learning sonata form structure on mozart's string quartets,''
  \emph{Transactions of the International Society for Music Information
  Retrieval (TISMIR)}, vol.~2, no.~1, 2019.

\bibitem{sears2018simulating}
D.~R. Sears, M.~T. Pearce, W.~E. Caplin, and S.~McAdams, ``Simulating melodic
  and harmonic expectations for tonal cadences using probabilistic models,''
  \emph{Journal of New Music Research}, vol.~47, no.~1, pp. 29--52, 2018.

\bibitem{hamilton2017inductive}
W.~Hamilton, Z.~Ying, and J.~Leskovec, ``Inductive representation learning on
  large graphs,'' \emph{Advances in neural information processing systems},
  vol.~30, 2017.

\bibitem{chawla2002smote}
N.~V. Chawla, K.~W. Bowyer, L.~O. Hall, and W.~P. Kegelmeyer, ``Smote:
  synthetic minority over-sampling technique,'' \emph{Journal of artificial
  intelligence research}, vol.~16, pp. 321--357, 2002.

\bibitem{li2019deepgcns}
G.~Li, M.~Muller, A.~Thabet, and B.~Ghanem, ``Deepgcns: Can gcns go as deep as
  cnns?'' in \emph{Proceedings of the IEEE/CVF international conference on
  computer vision}, 2019.

\bibitem{tonic_chord_2018}
\BIBentryALTinterwordspacing
``Bach: Prelude and fugue no.19 in a major, bwv 864 analysis,'' May 2018.
  [Online]. Available:
  \url{https://tonic-chord.com/bach-prelude-and-fugue-no-19-in-a-major-bwv-864-analysis/}
\BIBentrySTDinterwordspacing

\end{thebibliography}

\end{document}